
\magnification 1200

%

%
\font\eightrm=cmr8
\font\eighti=cmmi8
\font\eightsy=cmsy8
\font\eightbf=cmbx8
\font\eighttt=cmtt8
\font\eightit=cmti8
\font\eightsl=cmsl8
\font\sixrm=cmr6
\font\sixi=cmmi6
\font\sixsy=cmsy6
\font\sixbf=cmbx6
\catcode`@11
\newskip\ttglue
\font\grrm=cmbx10 scaled 1200

\def\eightpoint{\def\rm{\fam0\eightrm}
\textfont0=\eightrm \scriptfont0=\sixrm \scriptscriptfont0=\fiverm
\textfont1=\eighti \scriptfont1=\sixi \scriptscriptfont1=\fivei
\textfont2=\eightsy \scriptfont2=\sixsy \scriptscriptfont2=\fivesy
\textfont3=\tenex \scriptfont3=\tenex \scriptscriptfont3=\tenex
\textfont\itfam=\eightit \def\it{\fam\itfam\eightit}
\textfont\slfam=\eightsl \def\sl{\fam\slfam\eightsl}
\textfont\ttfam=\eighttt \def\tt{\fam\ttfam\eighttt}
\textfont\bffam=\eightbf
\scriptfont\bffam=\sixbf
\scriptscriptfont\bffam=\fivebf \def\bf{\fam\bffam\eightbf}
\tt \ttglue=.5em plus.25em minus.15em
\normalbaselineskip=6pt
\setbox\strutbox=\hbox{\vrule height7pt width0pt depth2pt}
\let\sc=\sixrm \let\big=\eightbig \normalbaselines\rm}
\newinsert\footins
\def\newfoot#1{\let\@sf\empty
  \ifhmode\edef\@sf{\spacefactor\the\spacefactor}\fi
  #1\@sf\vfootnote{#1}}
\def\vfootnote#1{\insert\footins\bgroup\eightpoint
  \interlinepenalty\interfootnotelinepenalty
  \splittopskip\ht\strutbox 
  \splitmaxdepth\dp\strutbox \floatingpenalty\@MM
  \leftskip\z@skip \rightskip\z@skip
  \textindent{#1}\footstrut\futurelet\next\fo@t}
\def\fo@t{\ifcat\bgroup\noexpand\next \let\next\f@@t
  \else\let\next\f@t\fi \next}
\def\f@@t{\bgroup\aftergroup\@foot\let\next}
\def\f@t#1{#1\@foot}
\def\@foot{\strut\egroup}
\def\footstrut{\vbox to\splittopskip{}}
\skip\footins=\bigskipamount 
\count\footins=1000 
\dimen\footins=8in 

\def\ref#1{$^{#1}$}
\def\flex{\raise 6pt\hbox{$\leftrightarrow $}\! \! \! \! \! \! }

\newbox\bigstrutbox
\setbox\bigstrutbox=\hbox{\vrule height10pt depth5pt width0pt}
\def\bigstrut{\relax\ifmmode\copy\bigstrutbox\else\unhcopy\bigstrutbox\fi}
\def\refer[#1/#2]{ \item{#1} {{#2}} }
\def\rev<#1/#2/#3/#4>{{\it #1\/} {\bf#2}, {#3}({#4})}
\def\boxit#1{\vbox{\hrule\hbox{\vrule\kern3pt
\vbox{\kern3pt#1\kern3pt}\kern3pt\vrule}\hrule}}

\def\2figure#1#2#3#4{\vbox{ \hrule width#1truecm \hbox{\vrule height#2truecm
\hskip #1truecm
\vrule height#2truecm }\hrule width#1truecm \hbox{\vrule\vbox{\hsize #1truecm
\baselineskip=10pt
\noindent\strut#3}\vrule}\hrule width#1truecm
\hbox{\vrule\vbox{\hsize #1truecm
\baselineskip=10pt
\noindent\strut#4}\vrule}\hrule width#1truecm  }}
\def\3figure#1#2#3#4#5{\vbox{ \hrule width#1truecm \hbox{\vrule height#2truecm
\hskip #1truecm
\vrule height#2truecm }\hrule width#1truecm \hbox{\vrule\vbox{\hsize #1truecm
\baselineskip=10pt
\noindent\strut#3}\vrule}\hrule width#1truecm
 \hbox{\vrule\vbox{\hsize #1truecm
\baselineskip=10pt
\noindent\strut#4}\vrule}
\hrule width#1truecm \hbox{\vrule\vbox{\hsize #1truecm
\baselineskip=10pt
\noindent\strut#5}\vrule}\hrule width#1truecm  }}

\def\sqr#1#2{{\vcenter{\hrule height.#2pt
   \hbox{\vrule width.#2pt height#1pt \kern#1pt
    \vrule width.#2pt}
    \hrule height.#2pt}}}


\def\smin{\,\raise 0.06em \hbox{${\scriptstyle \in}$}\,}
\def\smsubset{\,\raise 0.06em \hbox{${\scriptstyle \subset}$}\,}

\def\Natural{\hbox{\hskip 1.5pt\hbox to 0pt{\hskip -2pt I\hss}N}}

\def\Rational{\hbox{\hbox to 0pt{\hskip 2.7pt \vrule height 6.5pt
                                  depth -0.2pt width 0.8pt \hss}Q}}
\def\Real{\hbox{\hskip 1.5pt\hbox to 0pt{\hskip -2pt I\hss}R}}
\def\Complex{\hbox{\hbox to 0pt{\hskip 2.7pt \vrule height 6.5pt
                                  depth -0.2pt width 0.8pt \hss}C}}


1
\vfill\eject
\centerline {\grrm  Correlation Functions in Non Critical (Super) String
Theory}
\vskip 1cm
\centerline {\bf E. Abdalla$^1$, M.C.B. Abdalla$^2$,}

\centerline {\bf D. Dalmazi$^2$, Koji Harada$^{3}$}
\vskip .2cm

\centerline {$^1$Instituto de F\'\i sica, Univ. S\~ao Paulo, CP 20516,
S\~ao Paulo, Brazil}
\vskip .2cm

\centerline {$^2$Instituto de F\'\i sica Te\'orica, UNESP, Rua Pamplona 145,}

\centerline {CEP 01405, S\~ao Paulo, Brazil}
\vskip .2cm
\centerline {$^3$Department of Physics, Kyushu University, Fukuoka 812, Japan}
\vskip 2cm
\centerline {\bf Abstract}
\vskip .5cm

We consider the correlation functions of the tachyon vertex operator of the
super Liouville theory coupled to matter fields in the super Coulomb gas
formulation, on world sheets with spherical topology. After integrating over
the zero mode and assuming that the $s$ parameter takes an integer value,
we subsequently continue it to an arbitrary real number and
compute the correlators in a closed form.   We also included an arbitrary
number of screening charges  and,  as a result, after  renormalizing them,
as well as the external legs and the cosmological constant, the form of the
final amplitudes  do not modify. The result is remarkably parallel to the
bosonic  case. For completeness, we discussed the
calculation of bosonic correlators including arbitrary screening  charges.

\vfill \eject
\centerline {\bf  1- Introduction }
\vskip .50cm

Two dimensional gravity is not only a toy model for the theory of gravitation,
but also describes phenomena such as random surfaces and string theory away
from criticality\ref{1}. The discretized counterpart, namely matrix models,
proved to be an efficient means to obtain information, especially about non
critical string theory, while computing correlation functions\ref{2}.

In the continuum approach\ref{4}, in the conformal gauge, we have to face
Liouville theory\ref{5}.
However, although several important developments have been achieved\ref{6}, we
still lack some important points, in spite of much effort which has been spent.

In particular, it is difficult to calculate
correlation functions in a reliable way because  perturbation theory does not
apply. Recently, however, several authors\ref{7-14} succeeded in taming the
difficulties of the Liouville theory and computed exactly correlation functions
in the continuum approach to conformal fields coupled to two dimensional
gravity. The technique is based on the integration over the zero mode of
the Liouville field. The resulting amplitude is a function  of a
parameter $s$ which
depends on the central charge and on the external momenta. The
amplitudes can be computed when the above parameter is a non-negative
integer.
Later on, one analytically continues that parameter to  real (or complex)
values. The results for the correlation functions of the tachyon operator
thus obtained agree with the matrix model approach.\ref{2} General correlation
functions including arbitrary screening charges are also computed, and may be
useful for the purpose of studying fusion rules of minimal models coupled to
two dimensional gravity.\ref{25}

Opposite to the bosonic case supersymmetric matrix models are up to now
scarcely known. There are  few
papers in the literature concerning this approach\ref{3,25}, and it is
desirable to have
results from an alternative approach for comparison. For this reason,
several groups are studying the continuum approach to two
dimensional supergravity\ref{17-21}.
Our aim here is to investigate the supersymmetric Liouville theory.
We shall
compute supersymmetric correlation functions on world sheets with
spherical
topology in the Neveu-Schwarz sector, where the super-Liouville is
coupled
to superconformal matter with central charge $\hat c\le 1$, represented
as a super
Coulomb gas\ref{23,17}. The results are remarkable, and very parallel to the
bosonic case; after a redefinition of the cosmological constant,
and of the
primary superfields, the resulting  amplitudes have the same form as those of
the bosonic theory obtained by Di Francesco and Kutasov\ref{9}.  Our present
results generalize those presented in a recent paper\ref{19}, as well
as others recently obtained in the literature\ref{20,21,26}.

The paper is organized as follows. In section 2 we review some computations of
the bosonic correlators and generalize them to include an arbitrary number of
s.c. . In section 3 we calculate the $N$-point correlator of the Neveu-Schwarz
vertex operator in the $n=1$ two dimensional supergravity coupled to $N=1$
supermatter including an arbitrary number s.c..

 Our results include
the limiting cases $c=1$  ($\hat c=1$)  and in those situations some
physical conclusions can be drawn in the section 4: the amplitude factorizes,
and the expected
intermediate poles have zero residue, due to strong kinematic constraints. All
results are at least consistent with the matrix model approach, with possible
exception about the inclusion of screening charges, in the 3-point correlation
function. In the latter case, fusion rules deserve careful study.

\vskip 1cm
\penalty-200
\centerline {\bf 2- Bosonic Correlation Functions}
\nobreak
\vskip .5cm
\nobreak

Some $N$-point tachyon correlation function in Liouville theory coupled to
$c\le 1$ conformal matter were recently calculated by Di Francesco and
Kutasov\ref{9}. They worked in the DDK's
framework\ref{22} where the total action is given by:
$$
S={1\over 2\pi} \int d^2w \sqrt {\hat g}
\left[ \hat g^{ab}\partial_a\phi\partial _b\phi
-{Q\over 4}\hat R \phi
+2\mu e^{\alpha \phi} + {\hat g}^{ab} \partial _aX\partial _bX
+{i\alpha_0\over2}\hat R X\right]\quad , \eqno(2.1)
$$
here $\phi$ represents  the Liouville mode and $X$ is the matter field with
the central charge given by $c=1-12\alpha_0^2$. From the literature\ref{17}
we know that the constant $Q$ is determined
by imposing a vanishing total central charge, and is given by
$$ Q=2\sqrt {2+\alpha_0^2}.$$

The value of $\alpha$ is determined by requiring $e^{\alpha\phi}$ to be
a (1,1) conformal operator, yielding the equation  ${-\alpha\over 2}
(\alpha+Q)=1$, whose solutions are labeled by $\alpha _\pm$
$$
\alpha_\pm=-{Q\over 2}\pm \vert \alpha_0\vert \quad ,\quad
\alpha_+\alpha_-=2\eqno(2.2)
$$
the semiclassical limit $(c \to -\infty)$ fixes $\alpha =\alpha_+$.

The  gravitationally-dressed tachyon amplitudes are the objects we are
interested in:
$$
\langle T_{k_1}\cdots T_{k_N}\rangle =
\left\langle
\prod_{j=1}^N\int d^2z_je^{ik_jX(z_j)+\beta(k_j)\phi(z_j)}
\right\rangle
\eqno(2.3)
$$
where we fix the dressing parameter $\beta$ imposing  $e^{ik_jX+
\beta_j\phi}$ to be a  (1,1) conformal operator  and supposing the space-time
energy to be positive:
$$
E=\beta (k)+{Q\over 2} =\vert k_j-\alpha_0\vert
\quad .\eqno(2.4)
$$

In the calculation of the amplitudes $\langle T_{k_1}\cdots
T_{k_N}\rangle $ the main ingredient is the integration over the matter
$(X_0)$ and the Liouville $(\phi_0)$ zero modes. This is the so called
zero mode technique:  one splits\ref{4,5} both, the matter and
the Liouville fields as a sum of the zero mode $(X_0), (\phi_0)$ plus
fluctuations $(\tilde X), (\tilde \phi)$, where the
fluctuations are orthogonal to the zero mode. After such splitting we are left
with the following integrals:
$$
\eqalign{
\int _{-\infty}^{\infty}{\cal D}X_0e^{iX_0\left(\sum_{i=1}^Nk_i
-2\alpha_0\right) }
&=2\pi \delta\left( \sum_{i=1}^Nk_i-2\alpha_0 \right)\quad,\cr
\int _{-\infty}^{\infty}{\cal D}\phi_0
e^{i\phi_0\left(\sum_{j=1}^N\beta_j+Q\right)
-e^{\alpha_+\phi_0\left( {\mu \over \pi}\int d^2w
e^{\alpha_+\tilde\phi}\right)}
}& ={\Gamma (-s)\over -\alpha_+}
\left( {\mu\over \pi }\int d^2w e^{\alpha_+\tilde\phi}\right) ^s
\quad ,\cr }\eqno(2.5)
$$
where we have used that on the sphere ${1\over 8\pi} \int d^2 w \sqrt
{\hat g}\hat R=1$ and
$$
s=-{1\over \alpha_+}\left( \sum _{j=1}^N\beta_j+Q\right)
\quad .\eqno(2.6)
$$
We thus obtain for the amplitude
$$
\eqalign{
\langle T_{k_1}\cdots T_{k_N}\rangle &=2\pi \delta\left(
\sum_{j=1}^N k_j-2\alpha_0 \right) {\Gamma (-s)\over -\alpha_+}
\left( {\mu \over \pi}\right) ^s\cr
&\times \left\langle \prod_{j=1}^N\int d^2z_j e^{ik_j+\beta_j\phi(z_j)}
\left( \int d^2w e^{\alpha_+\phi}\right) ^s\right\rangle_0 \cr }
\eqno(2.7)
$$
where $\langle \cdots \rangle _0$ means that now the correlation
functions are
calculated as in the free theory $(\mu =0)$. The strategy to obtain
${\cal A}_N$
is to assume first that $s$ is a non-negative integer and to continue the
result to any real  $s$ at the end. Thus, using free propagators
$$
\langle X(w)X(z)\rangle_0=\langle\phi(w)\phi(z)\rangle_0
= \ln \vert w-z\vert ^{-2}
\eqno(2.8)
$$
and fixing the residual
$SL(2,\Complex )$ invariance of the conformal gauge on the sphere
by choosing $(z_1=0\, ,\, z_2=1\, ,\, z_3=\infty)$, the 3-point
function is written as:
$$
{\cal A}_3(k_1,k_2,k_3)={\Gamma (-s)\over -\alpha_+}
\left( {\mu \over \pi}\right) ^s\int \prod_{j=1}^sd^2w_j
\vert w_j\vert ^{2\alpha} \vert 1-w_j\vert ^{2\beta}
\prod_{i<j}^s\vert w_i-w_j\vert ^{4\rho}
\quad ,\eqno(2.9)
$$
where we have defined  $\alpha=-\alpha_+\beta_1\, ,\, \beta=-\alpha_+\beta_2
\, ,\,\rho=-\alpha_+^2/2$.   Choosing the
kinematics $k_1,k_3\ge \alpha_0 \, ,\, k_2<\alpha_0\le 0$ (notice that our
notation differs from Ref.[9] by the exchange of $k_2$ and  $k_3$.)  we can
eliminate $\beta$ using (2.4), (2.6) and the momentum conservation,
one can write the 3-point amplitude in a rather compact form
$$
{\cal A}_3=
[\mu \Delta(-\rho)]^s\prod_{j=1}^3\Delta \left( {1\over2}
(\beta_j^2-k_j^2)\right)\quad , \eqno(2.10)
$$
where $\Delta(x)=\Gamma(x)/\Gamma(1-x)$.
After redefinitions of the cosmological constant and of the external fields as
$$
\mu \to {\mu \over \Delta(-\rho)}\quad ,
\quad T_{k_j}\to {T_{k_j}\over \Delta
\left( {1\over 2}(\beta_j^2-k_j^2)\right)}\quad ,
\eqno(2.11)
$$
Di Francesco and Kutasov\ref{9} obtained  for the three-point function
$$
{\cal A}_3=\mu ^s \quad ,\eqno(2.12)
$$
which is also obtained in the matrix model approach.  In the following we shall
see that a similar expression holds for general $N$-point tachyon amplitudes
with an arbitrary number of screening charges.

The screening charges are introduced in the form of $n$ operators
$e^{id_+X}$ and $m$ operators $e^{id_-X}$,
with $d_\pm$ solutions of: ${1\over 2}d(d-2\alpha_0)=1\, ,\,
(d_+d_-=-\alpha_+\alpha_- =-2)$. Integrating over the zero-modes
again we get:
$$
\eqalign{
&\left\langle T_{k_1}T_{k_2}T_{k_3}
\left( {1\over n!}\prod _{i=1}^n\int d^2t_ie^{id_+X(t_i)}\right)
\left( {1\over m!}\prod _{i=1}^m\int d^2r_ie^{id_-X(r_i)}\right)
\right\rangle \cr
&= 2\pi \delta(\sum _{i=1}^3k_i+nd_++md_--2\alpha_0)
{\cal A}^{nm}_3(k_1,k_2,k_3)
\cr}\eqno(2.13)
$$
where the amplitude ${\cal A}^{nm}_3(k_1,k_2,k_3)$ is given by
the expression
$$\eqalign{
{\cal A}_3^{nm}(k_1,k_2,k_3)&
={\Gamma(-s)\over -\alpha_+}\left( {\mu \over \pi}\right) ^s
\prod _{i=1}^n\int d^2t_i\vert t_i\vert ^{2\tilde \alpha}
\vert 1-t_i\vert^{2\tilde \beta}
\prod _{i<j}^n\vert t_i-t_j\vert ^{4\tilde \rho}\cr
&\times \prod _{i=1}^m\int d^2r_i\vert r_i\vert^{2\tilde \alpha'}
\vert 1-r_i\vert^{2\tilde \beta'}
\prod_ {i<j}^m\vert r_i-r_j\vert ^{4\tilde\rho'}\cr
&\times \prod_{i=1}^{n}\prod_{j=1}^m\vert t_i-r_j\vert^{-4}
\prod_{i=1}^s\int d^2z_i\vert z_i\vert ^{2\alpha}
\vert 1-z_i\vert^{2\beta}\prod_{i<j}^s\vert z_i-z_j\vert ^{4\rho}\quad .\cr}
\eqno(2.14)
$$
The parameters  $\alpha,\beta$ and $\rho$ are defined as before,
and the remaining parameters  are
$$\eqalign{
\tilde  \alpha &=d_+k_1\quad ,\quad \tilde  \beta = d_+k_2\quad ,\quad
\tilde  \rho ={1\over 2}d_+^2 \cr
\tilde  \alpha '&=d_-k_1\quad ,\quad \tilde  \beta '= d_-k_2\quad
,\quad \tilde  \rho '={1\over 2} d_-^2 \quad .\cr}\eqno(2.15)
$$

Notice that the gravitational part of the amplitude (integrals over
$z_i$) is
the same as in the case  without screening charges. The
integrals over $t_i$ and $r_j$ (matter contributions) have been
calculated by Dotsenko and Fatteev\ref{16} (See their formula (B.10) in the
second paper of [16]); the result turns out to be
$$\eqalign{
{\cal A}_3^{nm}&=\! \left( {\mu \over \pi}\right)^s\Gamma(-s)
\Gamma(s\! +\! 1)
\pi^{s+n+m}\tilde \rho^{-4nm}
[\Delta(1\! -\! \tilde \rho)]^n[\Delta (1 \! -\!\tilde \rho']^m
\prod_{i=1}^m\Delta(i\tilde \rho'\! -\! n)\prod_{i=1}^n
\Delta(i\tilde \rho)\cr
&\times \prod_{i=0}^{m-1}
\Delta(1-n+\tilde \alpha' +i\tilde \rho')\Delta (1-n+\tilde \beta'
+i\tilde\rho' )
\Delta (-1+n-\tilde \alpha' -\tilde \beta' -(n-1+i)\tilde \rho')\cr
&\times \prod_{i=0}^{n-1}
\Delta(1+\tilde \alpha +i\tilde \rho)\Delta (1+\tilde \beta
+i\tilde\rho )
\Delta (-1+2m-\tilde \alpha -\tilde \beta -(n-1+i)\tilde \rho)\cr
&\times [\Delta (1-\rho)]^s\prod_{i=1}^s \Delta(i\rho)\prod_{i=0}^{s-1}
\Delta(1+ \alpha +i \rho)\Delta (1+\beta +i\rho )
\Delta (-\! 1 \! -\! \alpha\! -\! \beta\! -\! (s\! -\! 1\! +\! i)
\rho)\cr}\eqno(2.16)
$$

Assuming $\alpha_0<0$, and the same kinematics $(k_1\, ,\, k_3\ge \alpha_0
\, ,\, k_2 < \alpha_0)$ we can eliminate $\beta$ again, using (2.4),
(2.6) and momentum conservation. In this way,  we have
$$
\eqalign{
\tilde \alpha&= \alpha -2\rho \quad ,\quad \tilde \alpha'=-2+\tilde
\rho \alpha\cr
\beta &=-1-m-(s+n)\rho \quad ,\cr
\tilde \beta &= m-1+(s+n)\rho \quad ,\quad \tilde \beta ' = s+n+\rho
^{-1}(m-1)\cr
\tilde \rho &= -\rho \quad \quad ,\quad \tilde \rho ' =- \rho^{-1}\quad .\cr
}\eqno(2.17)
$$
Substituting in (2.16) we obtain the three point function with arbitrary
screenings
$$
\eqalign{
{\cal A}_3^{nm}&=\left( {\mu \over \pi}\right) ^s\Gamma(-s)\Gamma(s+1)
\pi^{s+n+m}(\tilde \rho)^{-4nm} \left[ \Delta(1+\rho^{-1})\right]^m
\left[ \Delta(1+\rho)\right]^n \cr
&\times\prod_{i=1}^m\Delta(i\rho^{-1}-n)\prod_{i=1}^n\Delta
(-i\rho ) \left[ \Delta (1-\rho)\right] ^s \prod_{i=1}^s\Delta(i\rho)\cr
&\times
\prod_{i=0}^{n-1}\Delta(m+(s+n-i)\rho)\prod_{i=0}^{s-1}\Delta(-m-
(s+n-i)\rho)\cr
&\times \prod_{i=0}^{m-1}\Delta(1+s+(m-1-i)\rho^{-1}) \cr
&\times \prod_{i=0}^{m-1}\Delta(-1-n+\rho^{-1}\alpha-i\rho^{-1})\Delta
(1-s-\rho^{-1}\alpha +i\rho^{-1})\cr
&\times \prod_{i=0}^{n-1}\Delta(1+\alpha-(i+2)\rho)
\Delta(m-\alpha - (s-1-i)\rho)\cr
&\times\prod_{i=0}^{s-1}\Delta(1+\alpha+i\rho)\Delta(m-\alpha
+ (n+1-i)\rho)\quad .\cr}\eqno(2.18)
$$

To get a simpler expression for ${\cal A}_3^{nm}$ we look for the term
$\Delta(\rho\!-\!\alpha) \Delta (\rho (s\!-\!n\!+\!1)\!+\!\alpha \!-\!m\!+\!1)
\times \break\Delta(-m\rho^{-1}\!-\!(s\!+\!n))$ which corresponds to
$\prod_{i=1}^3\Delta({1\over 2}
(\beta_i^2-k_i^2)) $. After algebraic manipulations we get
$$
{\cal A}_3^{nm}=\left[ \mu \Delta (-\rho)\right]^s\left[ -\pi \Delta
(\rho^{-1})\right]^m\left[ -\pi \Delta(\rho)\right]^n
\prod_{i=1}^3(-\pi)\Delta\left( {1\over2}(\beta_i^2-k_i^2)\right)
\quad .\eqno(2.19)
$$

This result has been also obtained by Di Francesco and
Kutasov\ref{9,20}, as well as Aoki and D'Hoker\ref{14}. Note that the factors
$\Delta(\rho)$ and $\Delta(\rho^{-1})$
 can be easily understood; the screening operators are renormalized like the
tachyon vertex operators $T_k$ with vanishing dressing $\beta(k)$.

Remembering the momentum conservation law:
$$\sum _{i=1}^3k_i=(1-n)d_++(1-m)d_-\quad ,\eqno(2.20)
$$
we can use (2.19) with $n=m=1$ and the formula below to obtain the partition
function ${\cal Z}$:
$$
{\partial ^3{\cal Z}\over \partial \mu^3}={\cal A}_3^{11}(k_i\to 0)\eqno(2.21)
$$
thus,
$$
{\cal Z}=\Delta(\rho)\Delta( \rho^{-1}){\rho^3[\mu
\Delta(-\rho)]^{1-\rho^{-1}}\over (\rho-1)(\rho+1)}\quad .\eqno(2.22)
$$
We  can also get the two-point function by taking, e.g., $k_3 \to 0$ in (2.19)
and (2.21):
$$
{\cal A}_3^{\tilde n\tilde m}(k_1,k_2,k_3 \to 0)={\partial \over \partial \mu
}{\cal A}_2^{\tilde n\tilde m}(k_1,k_2)\quad ,\eqno(2.23)
$$
where $(\tilde n,\tilde m)$ are fixed by (2.20) with $k_3=0$. Thus we finally
arrive at

$$
{\cal A}_2^{\tilde n\tilde m}(k_1,k_2)={\left[ \mu \Delta
(-\rho)\right]^{\rho^{-1}\left( {\alpha_+\over 2}\sum
_{i=1}^2\beta_i+\rho-1\right)}
\over \rho^{-1}\left( {\alpha_+\over 2}\sum _{i=1}^2\beta_i+\rho-1\right)}
\left[\Delta (\rho^{-1})\right]^{\tilde m}\left[\Delta(\rho)\right]^{\tilde
n}\prod_{i=1}^2\Delta\left( {(\beta_i^2-k_i^2)\over 2}\right)\eqno(2.24)
$$

We are able now to calculate ratios of correlation functions to compare with
other results in the literature. We have (in a generic kinematic region):
$$
R={{\cal A}_3^{nm}(k_1,k_2k_3){\cal Z}\over {\cal A}_2^{\tilde n_1,\tilde m_1}
(k_1,k_1){\cal A}_2^{\tilde n_2,\tilde m_2}(k_2,k_2){\cal A}_2^{\tilde n_3,
\tilde m_3}(k_3,k_3)}= {\prod _{i=1}^3\alpha_+\vert k_i-\alpha_0\vert\over
(\rho-1)(1+\rho)}\eqno(2.25)
$$

In the case of minimal models the momenta assume the following form:
$$
k_{r_ir'_i}={(1-r_i)\over 2}d_++{(1-r'_i)\over 2}d_-\quad .\eqno(2.26)
$$

Plugging back in (2.25) we find for the ratio $R$ the result
$$
R={\langle T_{r_1r'_1}T_{r_2r'_2}T_{r_3r'_3}\rangle ^2{\cal Z}\over
\langle T_{r_1r'_1}T_{r_1r'_1}\rangle\langle T_{r_2r'_2}T_{r_2r'_2}\rangle
\langle T_{r_3r'_3}T_{r_3r'_3}\rangle} = \left( {\alpha_+\over
2}\right)^3{\prod_{i=1}^3\vert r_id_++r'_id_-\vert \over
(\rho-1)(\rho+1)}\eqno(2.27)
$$
That is exactly the same result obtained by Dotsenko\ref{10} (in a given
kinematic region) which reduces to the result of Goulian and Li
for $r_i=r'_i$; all
these results agree with those obtained by the matrix model approach.

At this point the following remark is in order. After renormalizing the
screening charges:
$$
e^{id_+X}\to {e^{id_-X}\over \Delta (\rho)} \quad ,\quad
e^{id_-X}\to {e^{id_-X}\over \Delta (\rho^{-1})}\quad ,\eqno(2.28)
$$
and the tachyon operators $T_{k_i}$ as well as the cosmological constant $\mu$
as before (see (2.11)) we get the renormalized amplitude.
$$
 {\cal A}_3^{nm}= \mu ^s\quad .\eqno(2.29)
$$

Using the above result we would be able to exactly reproduce the ratio in
(2.27). Thus, the comparison of those ratios with matrix models is not a
precise test and only measures the scalling of the amplitudes w.r.t. the
cosmological constant. All  singularities contained in the $\Delta $ functions
cancel out in such ratios. If on the other hand compare,  the 3-point function
(2.19) directly with the matrix model result we would find a precise agreement
only for $c=1$ $(\alpha _0=0)$ where the amplitude can be written as a function
of the renormalized cosmological constant as:
$$
{\cal A}_3^{nm}\sim (\tilde \mu )^s\prod_{i=1}^3\Gamma (1- \sqrt 2\vert
k_i\vert )\quad .\eqno(2.30)
$$

For $c<1$ the amplitude obtained via matrix model are finite and the
singularities contained in (2.19)  are not observed. Neither do the fusion
rules for minimal models (see discussion in [27,28]) appear in (2.19),
although they can be seen via matrix models.

We now generalize the cases known in the literature for $N>3$ including
screening charges.
Repeating the zero-mode technique in the most general case of an $N$-point
function with arbitrary screening charges we have
$$
\eqalign{
{\cal A}_N^{nm}&=(-\pi)^3\left({\mu\over \pi}\right)^s\Gamma(-s)
\prod_{i=1}^N\int d^2z_i\prod_{j=1}^n\int {d^2t_j\over n!}\prod_{k=1}^m
\int {d^2r_k\over m!}\cr
&\times\prod_{l=1}^s\int d^2w_l\left\langle e^{ik_iX(z_i)}e^{id_+X(t_j)}
e^{id_-X(r_n)}\right\rangle_0\left\langle e^{\beta_i\phi(z_i)}
e^{\alpha_+\phi(w_l)}\right\rangle_0\quad ,\cr}
\eqno(2.31)
$$
where $s=-{1\over \alpha_+}(\sum_{i=1}^N\beta_i+Q)$ and
 a factor $\pi^3/\alpha_+$ has been absorbed in the measure of the path
integral. Fixing the $SL(2,\Complex)$ symmetry we get:
$$
\eqalign{
{\cal A}_N^{nm}&=(-\pi)^3\left( {\mu\over \pi}\right)^s \Gamma(-s)
I^{nm}_N\quad ,\cr
I^{nm}_N&=\int \prod_{j=4}^Nd^2z_j\vert z_j\vert^{2\alpha_j}
\vert 1-z_j\vert
^{2\beta_j}\prod_{i<j=4}^N\vert z_i-z_j\vert ^{4\rho_{ij}}\cr
&\times\int \prod_{i=1}^sd^2w_i\vert w_i\vert ^{2\alpha}\vert 1-w_i
\vert ^{2\beta}\prod_{i<j=1}^s\vert w_i-w_j\vert ^{4\rho}\prod_{i=1}^s
\prod_{j=4}^N\vert w_i-z_j\vert ^{2p_j}\cr
&\times \prod_{i=1}^nd^2t_i\vert t_i\vert ^{2\tilde \alpha}\vert 1-t_i\vert
^{2\tilde \beta}\prod_{i<j}^n\vert t_i-t_j\vert ^{4\tilde \rho}
\prod_{i=1}^n\prod_{j=4}^N\vert z_j-t_i\vert ^{2\tilde \alpha_j}\cr
&\times\int \prod_{i=1}^md^2r_i\vert r_i\vert ^{2\tilde\alpha'}\vert 1-r_i
\vert^{2\tilde\beta'}\prod_{i<j=1}^m\vert r_i-r_j\vert ^{4\tilde\rho'}
\prod_{i=1}^m\prod_{j=4}^N\vert z_j-r_i\vert ^{2\tilde\alpha'_j}\cr
&\times \prod_{i=1}^n\prod_{j=1}^m\vert t_i-r_j\vert ^{-4}\quad ,\cr}
\eqno(2.32)
$$
where
$$
\eqalign{
\alpha_j&=k_1k_j-\beta_1\beta_j\quad ,\quad \tilde \alpha_j=d_+k_j\cr
\beta_j&=k_2k_j-\beta_2\beta_j\quad ,\quad \tilde\alpha_j=d_-k_j\cr
\rho_{lj}&={1\over 2}(k_lk_j-\beta_l\beta_j)\quad ,p_j=-\alpha_+\beta_j\quad ,
\quad 4\le j,l\le N\quad .\cr}
\eqno(2.33)
$$

The integral above has been calculated by Di Francesco
and Kutasov\ref{9}, for the case $n=m=0$.
For arbitrary $n,m$ we shall use the same technique. Notice that translation
invariance
$$w_i\to 1-w_i\, ,\, z_i\to 1-z_i\, ,\,
t_i\to 1-t_i \, ,\, r_i\to 1-r_i$$
implies the symmetry relations
$$\alpha \leftrightarrow \beta\, ,\,
\alpha_j
\leftrightarrow \beta_j\, ,\, \tilde\alpha\leftrightarrow \tilde\beta\, ,\,
\tilde\alpha'\leftrightarrow \tilde
\beta'$$
so that after the elimination of the remaining parameters
as a function of $\alpha,\beta,p_j$ and $\rho$ $(j=4,5,\cdots, N-1)$,
$I_N^{nm}$ exhibits an $\alpha$-$\beta$ symmetry
$$
I_N^{nm}(\alpha,\beta,p_j,\rho)=I_N^{nm}(\beta,\alpha,p_j,\rho)
\quad .\eqno(2.34)
$$
Similarly by the inversion of all variables $w_i, z_i, t_i, r_i$ we have:
$$
I_N^{nm}(\alpha,\beta,p_j,\rho)=I_N^{nm}(-2-\alpha-\beta-2\rho(s-1)
-p_N-P, \beta,p_j,\rho)\eqno(2.35)
$$
where $P=\sum_{j=4}^{N-1}p_j$. Further information about $I_N^{nm}$
can be obtained in the limit $\alpha\to \infty $ (or $\beta \to \infty$),
by using a technique applied by Dotsenko and Fatteev\ref{16} in the case of
contour integrals, we found
$$
I_N^{nm}\approx \alpha^{2\beta +2\rho(s-N-n+3)+2P-2m}\eqno(2.36)
$$
where we have used the kinematics: $k_1,k_2,\cdots ,k_{N-1}\ge
\alpha_0\, ,\, k_N<\alpha_0 $
and assumed $\alpha_0<0$ To eliminate most of the parameters as
a function of $\alpha,\beta,p_j$ and $\rho$ we use (2.4), (2.6) and momentum
conservation. After such  elimination the symmetry (2.25) becomes:
$$
I_N^{nm}(\alpha,\beta,p_j,\rho)=I_N^{nm}(m-1-P-\alpha-\beta+\rho(N+n
-1-s),\beta,p_j,\rho)\eqno(2.37)
$$
Using Stirling's formula, it is not difficult to check that
the following Ansatz is consistent with (2.34), (2.36) and (2.37):
$$
\eqalign{
{\cal A}_N^{nm}&=f_N^{nm}(\rho,p_j)\Delta(\rho\!-\!\alpha)
\Delta(\rho\!-\!\beta)
\Delta(1\!-\! m\!+\!P\!+\!\alpha\!+\!\beta\!+\!\rho(s\!+\!2\!
-\!N\!-\!n)) \cr
{\cal A}_N^{nm}&=f_N^{nm}(\rho,p_j)\prod_{j=1}^3\Delta\left(
{1\over 2}(\beta_j^2-k_j^2)\right)\quad .\cr}\eqno(2.38)
$$

Now we can fix $f_N^{nm}(\rho,p_j)$ by using the 3-point function ${\cal A}_3^
{nm}$.
$$
{\cal A}_N^{nm}(k_1,k_2,k_j\to 0,k_N)=(-\pi)^{N-3}{\partial\over
\partial \mu} ^{N-3}{\cal A}_3^{nm}(k_1,k_2,k_N)\quad ,\quad
3\le j\le N-1\quad .\eqno(2.39)$$

Now using the result for ${\cal A}_3^{nm}$  we get:
$$
f_N^{nm}(\rho,p_j)\! =\! [-\pi\Delta(\rho^{-1})]^m [-\pi\Delta(\rho)]^n
\! \! \left( \! {\partial ^{N-3}\over \partial _\mu}
\mu^{s+N-3}\!\right) \! \!
[\Delta(-\rho)]^s\! \! \prod_{j=4}^N(-\pi)\Delta({1\over 2}(\beta_j^2
\! -\! k_j^2))\quad .\eqno(2.40)
$$
We finally return to  (2.38) and obtain
$$
\displaylines{
{\cal A}_N^{nm}=(s+N-3)(s+N-4)\cdots (s+1)\left[\mu\Delta (-\rho )\right]^s
\hfill\cr
\hfill \left[ -\pi\Delta (\rho^{-1})\right]^m\left[ -\pi\Delta(\rho )\right]^n
\prod_{j=1}^N (-\pi )\Delta ({1\over 2} (\beta_j^2-k_j^2))
 \quad ,\quad(2.41)\cr}
$$
therefore, redefining the screening operators, $T_{k_j}$ and $\mu$ as
before, we have:
$$
{\cal A}_N^{nm}={\partial^{N-3}\over\partial\mu}\mu^{s+N-3}\eqno(2.42)
$$
which is a remarkable result; however, it is valid only in the kinematic
region already mentioned.
In order to extend for general $k$, we can use the same
technique as used in [20]. Notice that the amplitude  (2.41) factorizes as in
the case without screening charges.

\vskip 1cm
\penalty-200
\centerline {\bf 3- Supersymmetric Correlators }
\vskip .5cm

\nobreak

In a recent paper\ref{19} we have calculated the 3- and 4-point NS
correlations
functions using DHK formulation\ref{17} of super Liouville theory
coupled to
superconformal matter on the sphere without screening charges.
The total action $S$
is given by the sum of the super Liouville action $S_{SL}$ and the matter piece
$ S_M $,
$$
\eqalign{
S_{SL}=&{1\over 4\pi}\int d^2{\bf z}\hat E\left( {1\over 2}
\hat D_\alpha
\Phi_{SL}\hat D^\alpha \Phi_{SL}-Q\hat Y\Phi_{SL}-4i\mu
e^{\alpha_+\Phi_{SL}}\right)\quad ,\cr
S_M =&{1\over 4\pi}\int d^2{\bf z}\hat E({1\over 2}\hat D_\alpha
\Phi_{M}\hat
 D^\alpha \Phi_{M} + 2i\alpha_0\hat Y\Phi_M)\quad ,\cr}\eqno(3.1)
$$
where $\Phi_{SL}, \Phi_M$ are super Liouville and matter superfields
respectively. The central charge of the matter sector is $c={3\over 2}\hat c
\, , \, (\hat c=1-8\alpha_0^2)$. Analogously to the bosonic case the parameters
$Q$ and $\alpha_\pm $ are given by (compare with (2.2))
$$Q=2\sqrt {1+\alpha_0^2}\quad ,\quad \alpha_\pm =-{Q\over 2} \pm
{1\over
2}\sqrt {Q^2-4}=-{Q\over 2}\pm \vert \alpha_0 \vert\quad ,\quad \alpha
_+\alpha _-=1 \quad .\eqno(3.2)$$
We shall call $\tilde \Psi _{NS}$ the gravitationally dressed primary
superfields, whose form  is given by
$\tilde  \Psi _{NS}({\bf z}_i,k_i)=d^2{\bf z}\hat E
e^{ik\Phi_M({\bf z})}e^{\beta(k)\Phi_{SL}({\bf z})}$,  where
$$ \beta (k)= -{Q\over 2} +\vert k-\alpha _0 \vert
\quad .\eqno(3.3)
$$

The calculation of the  three-point function of
the primary superfield $\tilde \Psi _{NS} $, involves the expression:
$$
\left\langle \prod_{i=1}^3\int \tilde  \Psi _{NS}({\bf
z}_i,k_i)\right\rangle \equiv \int [{\cal D}_{\hat E}\Phi_{SL}]
[{\cal D}_{\hat E} \Phi_M]
\prod _{i=1}^3\tilde  \Psi _{NS}({\bf z}_i,k_i)e^{-S}
\quad .\eqno(3.4)
$$

We closely follow the method  already used in the bosonic case.
After integrating over the bosonic zero modes we get
$$
\eqalign{&\left\langle \prod_{i=1}^3\int \tilde  \Psi _{NS}({\bf
z}_i,k_i)\right\rangle \equiv 2\pi \delta \left( \sum
_{i=1}^3k_i-2\alpha_0\right) {\cal A}_3(k_1,k_2,k_3)\quad ,\cr
&{\cal A}(k_1,k_2,k_3) = \Gamma (-s)({-\pi \over 2})^3({i\mu \over \pi
})^s\left\langle \!\int \! \prod _{i=1}^3d^2\tilde  {\bf
z}_ie^{ik_i\Phi_M(\tilde  {\bf z}_i)}e^{\beta_i\Phi_{SL}(\tilde  {\bf
z}_i)}\left(\! \int \! d^2{\bf z}e^{\alpha_+\Phi_{SL}({\bf z})}
\! \right)
^s\right\rangle_0  \cr }\eqno(3.5)
$$
where $\langle \cdots \rangle _0$ denotes again the expectation value
evaluated in
the free theory $(\mu =0)$  and we have absorbed the factor
$[\alpha _+(-\pi/2)^3]^{-1}$ into the normalization
of the path integral. the parameter $s$ is
defined as in the bosonic case (see (2.6)).

For $s$ non-negative integer, after fixing the $\widehat{SL}_2$ gauge,\ref{30}
 $\tilde z_1=0\, ,\, \tilde z_2=1\, ,\, \tilde z_3=\infty\, ,\,
\tilde\theta_2=\tilde\theta_3=0\, ,\, \tilde\theta_1=\theta$, in  components
($\Phi_{SL}=\phi +\theta\psi +\bar \theta \bar \psi$) (the integral above is
the
supersymmetric generalization of (B.9) of Ref.[16]) we have
$$
\eqalign{&{\cal A}(k_1,k_2,k_3) = \Gamma (-s)({-\pi \over 2})^3
({i\alpha_+^2\mu
\over \pi})^s\beta_1^2\cr
\times &\int \prod _{i=1}^sd^2z_i
\prod_{i=1}^s\vert z_i\vert ^{-2\alpha_+\beta_1}\vert
1\! -\!  z_i\vert ^{-2\alpha_+\beta_2}\prod_{i<j}^s\vert z_i\! -\! z_j
\vert ^{-2\alpha_+^2}\langle \overline \psi \psi (0)\overline \psi
\psi(z_1)\cdots \overline \psi \psi (z_s)\rangle_0\quad .}\eqno(3.6)
$$
Observe that this is non-vanishing only for
$s$ odd ($s=2l+1$). One
may
evaluate $\langle \overline \psi \cdots \overline \psi \rangle_0$
and $\langle
\psi \cdots \psi\rangle_0$ independently, since the rest of the
integrand is
symmetric, one may write the result in a simple form by relabelling
coordinates:
$$
\eqalign{{\cal A}_3(k_1,k_2,k_3)& = \Gamma (-s)({-\pi \over 2})^3{1\over
\alpha_+^2} ({i\alpha_+^2\mu \over \pi})^s
\alpha^2(-1)^{{s+1\over 2}}s!!\cr
\times &\int \prod _{i=1}^sd^2z_i\prod_{i=1}^s
\vert z_i\vert ^{2\alpha}\vert
1-z_i\vert ^{2\beta} \prod_{i<j}^s\vert z_i-z_j\vert^{4\rho}
\prod_{i=1}^{s-1\over 2}\vert z_{2i-1}-z_{2i}\vert ^{-2}
\vert z_s\vert^{-2}\quad  \cr}$$

Redefining the variables as $ z_s\equiv w\, ,\quad  z_{2i-1}
\equiv \zeta_i \, $ and $ z_{2i} \equiv \eta_i$  we have:
$$
{\cal A}(k_1,k_2,k _3)=-i{-\pi\over 2}^3\Gamma(-s)\Gamma(s+1){1\over \alpha_
+^2}\left(
{\alpha_+^2\mu \over \pi }\right)^s I^l(\alpha,\beta;\rho)\quad ,\eqno(3.7)
$$
where
$$
\eqalign{
&I^l(\alpha,\beta;\rho)=
{1\over 2^ll!}\alpha^2\int d^2w\prod_{i=1}^ld^2\zeta_id^2\eta_i
\vert w\vert^{2\alpha-2} \vert 1-w\vert ^{2\beta}
\prod_{i=1}^l\vert w-\zeta_i\vert ^{4\rho}
\vert w-\eta_i\vert ^{4\rho}\cr
&\times \prod_{i=1}^l\vert \zeta_i\vert ^{2\alpha}
\vert 1-\zeta_i\vert^{2\beta} \vert 1-\eta_i \vert ^{2\beta}
\prod_{i,j}^l\vert
\zeta_i-\eta_j\vert^{4\rho}
\prod _{i<j}^l\vert \zeta_i-\zeta_j\vert ^{4\rho}\vert
\eta_i -\eta_j\vert ^{4\rho}
\prod_{i=1}^l\vert \zeta_i-\eta_i\vert ^{-2}\quad ,\cr
}\eqno(3.8)
$$
and $\alpha,\beta, \rho $ are defined as before.
In  ref.[19] we calculated $I^l$ in detail by using the symmetries
$I^l(\alpha,\beta;\rho)=I^l(\beta,\alpha;\rho)$, $
I^l(\alpha,\beta;\rho)=I^l(-1-\alpha-\beta-4l\rho,\beta;\rho)$ and looking at
its large $\alpha$ behavior we obtained:
$$
\eqalign{I^l(\alpha,\beta;\rho)=&-{\pi^{2l+1}\over 2^{2l}}
\left[ \Delta \left({1\over 2}-\rho \right) \right]^{2l+1}
\prod_{i=1}^l\Delta (2i\rho)
\prod_{i=1}^{l}\Delta  \left( {1\over 2}+(2i+1) \rho \right) \cr
&\times\prod_{i=0}^l\Delta( 1+\alpha +2i\rho) \Delta( 1+\beta+2i
\rho) \Delta( -\alpha-\beta+(2i-4l)\rho)\cr
&\times\prod_{i=1}^l\Delta( {1\over 2}+\alpha+(2i-1
)\rho) \Delta( {1\over 2}+\!\beta\!+\!(2i\!-\!1)
\rho) \Delta( -{1\over 2}\!-\!\alpha\!-\!\beta\!+\!(2i\! -\!4l\!-\!1)\rho)
\cr}\eqno(3.9)
$$

 We can choose,  $k_1,k_3\ge\alpha_0, k_2\le
\alpha_0$. We proceed now as in the bosonic case, obtaining for the parameter
$\beta$,
$$
\beta=\cases {\rho (1-s)\; \; (\alpha_0>0)\cr
-{1\over 2}-\rho s \; \; (\alpha_0<0)\quad .\cr}\eqno(3.10)
$$

Now we are ready to write down the amplitude. For $\alpha_0<0$ we have the
non-trivial amplitude:
$$\eqalignno{
{\cal A}(k_1,k_2,k_3)&= ({-i\pi\over 2})^3
\left[ {\mu\over 2}\Delta\left({1\over 2}-\rho\right) \right]^s
\Delta\left( {1\over 2}-{s\over 2}
\right)\Delta \left( 1+\alpha-(s-1)\rho\right) \Delta\left(
{1\over 2}-\alpha+\rho\right)\cr
&=\left[ {\mu\over 2}\Delta\left( {1\over 2}-\rho\right)\right]^s
\prod_{j=1}^3(-{i\pi\over 2})\Delta\left( {1\over 2}
[1+\beta_j^2-k_j^2]\right) &(3.11)\cr }
$$

In the case $\hat c=1$, we obtain for  the external legs, renormalization
factors of the form
$\Delta (1-\vert k_i\vert )$, which should be compared to the bosonic case
(2.30); it
permits as well comparison to super matrix model as well, whenever those are
available.

 By redefining the cosmological constant and the primary superfield
$\tilde \Psi_{NS}$
$$
\mu  \to  {2\over \Delta \left( {1\over 2}-\rho\right) }\mu
\quad,\quad
\tilde \Psi_{NS}(k_j)\to {1\over (-{i\over 2}\pi)\Delta\left( {1\over
2}[1+\beta_j^2-k_j^2]\right)}\tilde \Psi_{NS}(k_j)\quad , \eqno(3.12)
$$
we get
$$
{\cal A}_3(k_1,k_2,k_3)=\mu ^s\quad . \eqno(3.13)
$$
As in the bosonic case we have a remarkably simple result. The only
differences with respect to the bosonic case are in  the details of the
renormalization factors.
Compare (3.12) with (2.11). Note that the singular point at
the renormalization
of the cosmological constant is $\rho=-1$ in the bosonic case, which
corresponds to $c=1$, and $\rho=-{1\over 2}$ in the
supersymmetric case, corresponding to $\hat c=1$ or $c=3/2$.

We shall now generalize the above result to the case which includes
screening
charges in the supermatter sector. We consider $n$
charges $e^{id_+\Phi_M}$
and $m$ charges $e^{id_-\Phi_M}$, where $d_\pm $ are solutions of
the equation
 ${1\over 2}d(d-2\alpha_0)={1\over 2}$. After integrating over the
matter and
Liouville zero modes we get
$$
\eqalignno{
&\left\langle\prod_{i=1}^3\int \tilde  \Psi_{NS}(\tilde  {\bf z}_i,k_i)
\prod_{i=1}^n\int {d^2{\bf t}_i\over n!}e^{id_+\Phi_M({\bf t}_i)}
\prod_{i=1}^m\int
{d^2{\bf r}_i\over m!}e^{id_-\Phi_M({\bf r}_i)}  \right \rangle \cr
&\equiv 2\pi \delta \left( \sum _{i=1}^3k_i+nd_+
+md_- -2\alpha_0\right)
{\cal A}_3^{nm}(k_1,k_2,k_3)\cr
&{\cal A}_3^{nm}(k_1,k_2,k_3)=\Gamma(-s)({-\pi\over 2})^3({i\mu
\over \pi})^s\! \left\langle
\prod_{i=1}^n\int {d^2{\bf t}_i\over n!}e^{id_+\Phi_M({\bf
t}_i)} \prod_{i=1}^m\int {d^2{\bf r}_i\over m!}e^{id_-\Phi_M({\bf r}_i)}
\right.\cr
&\left.\times \int \prod_{i=1}^3d^2{\bf \tilde z}_i
e^{ik_i\Phi_M( \tilde {\bf
z}_i)}e^{\beta_i\Phi_{SL}({\bf \tilde z}_i)}
\left(\! \int d^2{\bf z} e^{\alpha_+
\Phi_{SL} ({\bf z})}\right)^s\!\right\rangle_0 \! ,&(3.14)\cr}
$$

Integrating over the Grasmann variables and fixing the
$\widehat {SL(2)}$ symmetry
as before $(\tilde  z_1=0\, ,\, \tilde z_2=1 \, ,\, \tilde  z_3=\infty\, ,\,
\tilde\theta_1=\theta\, ,\, \tilde  \theta_2=\tilde  \theta_3=0)$ we obtain
(using $d_+d_-=-\alpha_+\alpha_-=-1$)
$$\eqalignno{
{\cal A}_3^{nm}(k_1,k_2,k_3)
&=\Gamma(-s)\left( {-\pi\over 2}\right) ^3\left( {i\mu \alpha_+^2\over
\pi}\right) ^s{(-d_+^2)^n\over n!}{(-d_-^2)^m\over m!}\cr
&\times\prod _{i=1}^n\int d^2t_i\vert t_i\vert ^{-2d_+k_1}
\vert 1-t_i\vert^{-d_+k_2}\prod _{i<j}^n
\vert t_i-t_j\vert ^{2d_+^2}\cr
&\times \prod _{i=1}^m\int d^2r_i\vert r_i\vert^{-2d_-k_1}
\vert 1-r_i\vert^{-2d_-k_2}
\prod_ {i<j}^m\vert r_i-r_j\vert ^{2d_-^2}\prod_{i=1}^{n}
\prod_{j=1}^m\vert
t_i-r_j\vert^{-2}\cr
&\times \prod_{i=1}^s\int d^2z_i\vert z_i\vert ^{-2\alpha_+\beta_1}\vert
1-z_i\vert^{-2\alpha_+\beta_2}\prod_{i<j}^s
\vert z_i-z_j\vert ^{-2\alpha_+^2}\cr
&\times \left\langle (\beta_1^2\overline \psi\psi(0)-k_1^2\overline \xi
\xi(0))\prod _{i=1}^n\overline \xi \xi (t_i)
\prod _{i=1}^m\overline \xi \xi
(r_i) \prod_{i=1}^s\overline \psi \psi (z_i)\right\rangle_0 \quad .&(3.15)\cr }
$$

Since the vacuum expectation value of an odd number of
$\overline \psi \psi $
(or $\overline \xi \xi$ ) operators is zero we have only two
non-trivial cases:
in the first case $n+m=$ odd, $s=$ even and in the second one
$n+m=$ even, $s=$ odd. Thus we have (see also [21] for comparison)
$$\eqalign{
{\cal A}_3^{nm}(k_1,k_2,k_3)
&=\Gamma(-s)\left( {-\pi\over 2}\right) ^3
\left( {i\mu \alpha_+^2\over
\pi}\right) ^s{(-d_+^2)^n\over n!}{(-d_-^2)^m\over m!}\cr
&\times \cases {I^{nm}_M(\tilde  \alpha,
\tilde \beta;\overline \rho)\,\times \,
I^s_G(\alpha,\beta;\rho)\, ,\, n+m={\rm even }\, ,\,
s={\rm odd}\cr
{}\cr
J^{nm}_M(\tilde  \alpha, \tilde \beta;\overline \rho)\,\times \,
J^s_G(\alpha,\beta;\rho)\, ,\, n+m={\rm odd }\, ,\, s={\rm
even}\cr}}\eqno(3.16)
$$
where
$$
\eqalignno{
I^{nm}_M(\tilde  \alpha,\tilde  \beta;\overline  \rho)& =
\prod _{i=1}^n\int d^2t_i\vert t_i\vert ^{2\tilde  \alpha}\vert 1-
t_i\vert^{2\tilde  \beta}\prod _{i<j}^n\vert t_i-t_j\vert ^{2\overline
\rho}\cr
&\times \prod _{i=1}^m\int d^2r_i\vert r_i\vert^{2\tilde \alpha'}
\vert 1-
r_i\vert^{2\tilde \beta'}\prod_ {i<j}^m\vert r_i-r_j\vert ^{2\overline
\rho'}\prod_{i=1}^{n}\prod_{j=1}^m\vert t_i-r_j\vert^{-2}\cr
&\times \left\langle \prod _{i=1}^n\overline \xi \xi (t_i)
\prod_{i=1}^m\overline \xi \xi (r_i)
\right\rangle_0\quad , &(3.17)\cr}
$$
 $$
I^s_G(\alpha,\beta;\rho)=\alpha^2\int \prod _{i=1}^sd^2z_i
\prod_{i=1}^s\vert
z_i\vert ^{2\alpha}\vert 1-z_i\vert ^{2\beta}
\prod_{i<j}^s\vert z_i-z_j\vert^{4\rho}
\left\langle\overline \psi\psi (0)
\prod_{i=1}^s\overline \psi \psi
(z_i) \right\rangle _0\eqno(3.18)
$$
with $\overline\rho =d_+^2,\overline\rho '=d_-^2 $
and $\alpha ,\beta ,\rho, \tilde \alpha, \tilde \alpha', \tilde \beta, \tilde
\beta' $  defined as before. Note that
$I^{nm}_M$ is the
supersymmetric generalization of (B.10) of the second reference in [16].
The integral $J^{nm}_M$
differs from $I^{nm}_M$ by the introduction of a factor
$\overline \xi \xi(0)$
and $J^s_G$ can be obtained from $I^s_G$  by dropping
$\overline \psi\psi(0)$.
Henceforth we assume,
for simplicity, $n+m=$ even, $s=$ odd. We will work out explicitly
only the
case $n,m$ even. However, the final result for the amplitude does
not depend on which case we choose. In ref. [24]  we have calculated
$I_M^{nm}$ for $n$ and $m$ even and we get:
$$
\eqalign{
&I^{nm}_M(\tilde  \alpha,\tilde \beta;\overline \rho)=
(-)^{{n+m\over 2}}{\pi^{n+m}\over 2^{n+m}}n!m!
\left( -{\overline\rho\over2}\right) ^{-2nm}
\left[\Delta\left( {1\over 2}-{\overline \rho\over 2}\right)\right] ^n
\left[\Delta\left( {1\over 2}-{\overline\rho'\over 2}\right)\right] ^m
\cr
&\times \prod _1^{n\over 2}\Delta (i\overline\rho)\Delta\left( {1\over
2}+\overline\rho\left( i-{1\over 2}\right) \right) \prod _1^{m\over 2}
\Delta (i\overline\rho' -{n\over 2})\Delta\left( {1\over 2}-{n\over 2}
-\overline\rho'\left( i-{1\over 2}\right) \right) \cr
&\times \prod _{i=0}^{{n\over 2}-1}\Delta (1+\tilde \alpha + i\overline
\rho )\Delta(1+\tilde  \beta +i\overline\rho)\Delta(m-\tilde  \alpha
-\tilde  \beta +\overline\rho (i-n+1))\cr
&\times \prod _{i=1}^{{n\over 2}}\Delta ({1\over 2}\! +\! \tilde \alpha
\!+\! (i \! - \! {1\over 2} )\overline
\rho )\Delta({1\over 2}\! +\! \tilde  \beta \! + \! (i\! -\! {1\over 2})
\overline\rho)
\Delta(-{1\over 2}\! -\! \tilde  \alpha \! +\! m \! -\tilde  \beta \! +
\! \overline\rho (i\! -\! n\! +\! {1\over 2}))\cr
&\times \prod _{i=0}^{{m\over 2}-1}\Delta (1+\tilde \alpha' - {n\over 2}
+ i\overline\rho ')\Delta(1 - {n\over 2} + \tilde  \beta' +i\overline
\rho' )\Delta( {n\over 2} -\tilde  \alpha' -\tilde  \beta '+\overline
\rho' (i-m+1))\cr
&\times \prod _{i=1}^{{m\over 2}}\Delta ({1\over 2}\!
-\!  {n\over 2}\!  +\!
\tilde  \alpha '\! +\! (i\! -\! {1\over 2} )\overline
\rho' )\Delta({1\over 2}\! -\! {n\over 2}\! +\! \tilde  \beta' \! +\!
(i\! - \! {1\over 2})\overline\rho')
\Delta(\! -{1\over 2}\! + {n\over 2}\! -\! \tilde  \alpha'\! -\! \tilde
\beta'\! +\! \overline\rho'(i\! -\! m \! + \! {1\over 2}))\cr}\eqno(3.19)
$$
In the case where $s=2l+1$ the gravitational contribution to
${\cal A}_3^{nm}(k_1,k_2,k_3) $, i.e, $I^s_G$ is just the same
as in the case
without screening charges, thus from the last section we have the
supersymmetric generalization of (B.9) of ref.[16]:
$$
\eqalign{
I^s_G & = (-)^{{s-1\over 2}}{\pi^s\over 2^{s-1}}s!
\left[ \Delta \left(
{1\over 2} -\rho\right) \right] ^s
\prod_{i=1}^{{s-1\over 2}}\Delta (2i\rho)
\prod_{i=0}^{{s-1\over 2}}\Delta ( {1\over 2} + (2i+1)\rho) \cr
&\times \prod _{i=0}^{{s-1\over 2}}\Delta (1+\tilde \alpha + 2i\tilde
\rho )\Delta(1+\tilde  \beta +2i\overline\rho)\Delta(-\tilde  \alpha
-\tilde  \beta +2\overline\rho (i-s+1))\cr
&\times \prod _{i=0}^{{s-1\over 2}}\Delta ({1\over 2}+\tilde \alpha +
(2i-1)\overline\rho )\Delta({1\over 2} +\tilde  \beta +(2i - 1)\overline\rho)
\Delta(-{1\over 2}-\tilde  \alpha -\tilde  \beta +\overline\rho
(2i-2s+1))\cr}\eqno(3.20)
$$

To obtain ${\cal A}_3^{nm}(k_1,k_2,k_3) $ (see (3.16)) we have to
calculate
$I^{nm}\times I^s_G$. Using the same kinematics as
in the case without screening
charges, and considering that $\alpha_0<0$   it is easy to deduce:
$$
\eqalign{
\tilde  \alpha&=\alpha-2\rho\quad ,\quad \tilde  \alpha'
=-1+{\rho^{-1}\alpha\over 2} \cr
\overline\rho&=-2\rho \quad \quad , \quad \overline\rho'= -{\rho^{-1}
\over 2} \cr
\beta &= -{1\over 2}-{m\over 2}-(n+s)\rho\cr
\tilde  \beta &=-\beta-1=(n+1)\rho +{m\over 2}-{1\over 2}\cr
\tilde  \beta'&={(n+s)\over 2}+{\rho^{-1}\over 4}(m-1)\quad .\cr}
\eqno(3.21)
$$
Substituting in (3.19) and (3.20) and using (3.16)
we obtain a very involved expression (see also [26,21]):
$$
\eqalign{
&{\cal A}_3^{nm}(k_1,k_2,k_3)\cr
&= \Gamma(-s)\left( {-\pi\over 2}\right) ^3\left( {i\mu
\over \pi}\right)^s\alpha_+^{2(s-1)}(2\rho)^{n-m}(-)^{n+m+s-1\over
2}{\pi^{s+n+m} \over 2^{m+n+s-1}}\rho^{-2mn}s! \cr
&\times \left[ \Delta({1\over 2}+\rho)\right]^n
\left[ \Delta({1\over
2}+{\rho^{-1}\over 4})\right]^m \cr
&\times \prod_{i=1}^{n\over 2}\Delta(-2i\rho)\Delta ({1\over 2}+\rho
(1-2i))\prod _{i=1}^{m\over 2} \Delta (-{n\over 2}-{i\rho^{-1}\over 2})
\Delta
\left( {1\over 2} -{n\over 2} +\left( {1\over 4}-{i\over 2}\right)
\rho^{-1}\right) \cr
&\times \prod _{i=0}^{{n\over 2}-1}\Delta \left( {1\over 2}
+{m\over 2}+\rho
(n+s-2i)\right) \prod_{i=1}^{n\over 2}\Delta\left( {m\over
2}+\rho(n+s+1-2i)\right) \cr
&\times \prod _{i=0}^{{m\over 2}-1}\Delta \left( 1+{s\over 2}
-{\rho^{-1}\over2} (i+{1\over 2}-{n\over 2})\right)
\prod_{i=1}^{m\over 2}\Delta\left( {1\over
2}+ {s\over 2}- {\rho^{-1}\over 2}(i-{m\over 2})\right) \cr
&\times \left[ \Delta ({1\over 2}-\rho)\right] ^s
\prod _{i=1}^{{s-1\over
2}}\Delta (2i\rho)\prod_{i=0}^{{s-1\over 2}}\Delta \left( {1\over
2}+(2i+1)\rho\right) \cr
&\times\prod_{i=0}^{{s-1\over 2}}\Delta ({1\over 2}-{m\over
2}+(2i-n-s)\rho)\prod _{i=1}^{{s-1\over 2}}\Delta(-{m\over
2}+(2i-1-n-s)\rho)\cr
&\times \prod_{i=1}^{{n\over 2}}\Delta({1\over
2}+\alpha-\rho(2i+1))\Delta({1\over 2}+{m\over 2} -\alpha
-\rho(s-n-2+2i))\cr
&\times \prod_{i=0}^{{s-1\over 2}}\Delta({m\over 2}+{1\over 2}
-\alpha+(2i-s+n+2)\rho)
\prod_{i=1}^{{s-1\over 2}}\Delta({1\over 2}+\alpha+(2i-1)\rho)
\cr
&\times \prod_{i=1}^{{n\over 2}}\Delta({m\over
2}-\alpha-\rho(2i-n+s-1))\Delta(1+\alpha-2i\rho)\cr
&\times \prod_{i=0}^{{s-1\over 2}}\Delta(1+\alpha+
2i\rho)\prod_{i=1}^{{s-1\over 2}}\Delta({m\over 2}-\alpha
+(2i-s+n+1)\rho) \cr
&\times \prod_{i=1}^{{m\over 2}}\Delta({1\over
2}-{s\over 2}-{\rho^{-1}\over 2}(i+\alpha -{m\over 2}))
\Delta(-{n\over 2} -
{\rho^{-1}\over 2}(i-1-\alpha))\cr
&\times \prod_{i=1}^{{m\over 2}}\Delta(-{n\over 2}-{1\over 2}
-{\rho^{-1}
\over 2} (i-\alpha-{1\over 2}))\Delta(1-{s\over 2}-{\rho^{-1}\over
2}(i-{(m+1)\over 2}+\alpha))\quad . \cr }\eqno(3.22)
$$

In order to obtain a simple expression for the amplitude we have
to combine in
each term  the matter and the gravitational parts as in the bosonic
case. The
calculation is more complicate now, but we finally get
$$
\eqalignno{
{\cal A}_3^{nm}(k_1,k_2,k_3)&= \left( -{\pi\over 2}\right)^3
\left[ {\mu \over 2} \Delta ({1\over 2}-\rho)\right]^s
\left[-{i\pi\over 2}\Delta({1\over 2}+{\rho^{-1}\over 4})\right] ^m
\left[-{i\pi\over 2}\Delta({1\over 2}+\rho)\right] ^n\cr
&\times\Delta\left(\rho-\alpha+{1\over 2}\right)\Delta\left({1\over
2}-{n+s\over 2} -{m\rho^{-1}\over 4}\right) \Delta(1-{m\over 2}+\alpha
+(s-n-1)\rho)\cr
&=\left[ {\mu \over 2} \Delta \left( {1\over 2}-\rho\right) \right]^s
\left[-{i\pi\over 2}\Delta\left( {1\over 2}+{\rho^{-1}\over 4}\right)
\right] ^m
\left[-{i\pi\over 2}\Delta\left( {1\over 2}+\rho\right) \right] ^n
\cr
&\times\prod_{i=1}^3 \left( -{i\pi\over 2}\right) \Delta \left(
{1\over 2}
+{1\over 2}(\beta_i^2-k_i^2)\right) \quad . &(3.23)\cr}
$$
Therefore after redefining the cosmological constant, the NS
operators and the screening charges
$$\eqalignno{
e^{id_+\Phi_M(t_i)}&\to
\left[ -{i\pi\over 2}\Delta\left( {1\over2}+\rho\right)
\right]^{-1}e^{id_+\Phi_M(t_i)} & (3.24a)\cr
e^{id_-\Phi_M(t_i)}&\to \left[ -{i\pi\over 2}\Delta\left( {1\over
2}+{\rho^{-1}\over 4} \right) \right]^{-1} e^{id_-\Phi_M(t_i)}
&(3.24b)\cr
\Psi_{NS}&\to \left[ -{i\pi\over 2} \Delta \left( {1\over 2}
+{1\over 2}(\beta_i^2-k_i^2)\right) \right]^{-1}\Psi_{NS}
& (3.24c)\cr
\mu &\to \left[ {1\over 2}\Delta \left( {1\over 2}-\rho\right)
\right]^{-1} \mu                             \quad .
& (3.24d)\cr }
$$
we obtain the very simple result:
$$
{\cal A}_3^{nm}(k_1,k_2,k_3)=\mu ^s\quad .\eqno(3.25)
$$
In view of the complexity of (3.22), the simplicity of the result is
remarkable.

As in the bosonic case we can calculate ratios of correlation functions either
using (3.23) or (3.25). We  obtain in a generic kinematic region:
$$
\eqalignno{
R&={\langle \Psi _{NS}(k_{r_1r'_1})\Psi
_{NS}(k_{r_2r'_2})\Psi_{NS}(k_{r_3r'_3})
 \rangle^2{\cal Z}\over\prod_{j=1}^3\langle \Psi _{NS}(k_{r_jr'_j})\Psi _{NS}
(k_{r_jr'_j})\rangle } &(3.26)\cr
R&= (2\alpha_+)^3{\prod_{i=1}^3\vert r_id_-+r'_id_+\vert \over
(2\rho-1)(2\rho+1)}\quad .&(3.27)\cr}
$$
Compare with (2.27).

The above result agrees with other results\ref{26,21} simultaneously obtained
in the literature. Although the the continuations to non integer values of
$s$ used in [26] and [21] are not the same and, in principle, do not correspond
to the procedure used here, the physical results seem to be independent of such
details.

We now show that it is possible to obtain a simple result for the most
general  case of a $N$-point amplitude with an arbitrary number of
screening charges (${\cal A}^{nm}_N$). In that general case to compute the
amplitudes we have to calculate the following integral
$$\eqalign{
{\cal A}^{nm}_N(k_1,&\cdots ,k_N)={\Gamma(-s)\over -\alpha_+}\left(
{i\mu \over
\pi}\right)^s\left\langle \prod_{i=1}^N\int d^2\tilde  {\bf
z}_ie^{ik_i\Phi_M(\tilde  {\bf z}_i)+\beta_i\Phi_{SL}(\tilde  {\bf
z}_i)}\right. \cr
&\left.\times \prod_{i=1}^n\int d^2{\bf t}_ie^{id_\Phi_M(\tilde
{\bf t}_i)}\prod_{j=1}^m\int d^2{\bf r}_je^{id_-\Phi_M
(\tilde  {\bf r}_j)}
\prod_{j=1}^s\int d^2{\bf z}_je^{\alpha_+\Phi_{SL}(\tilde  {\bf z}_i)}
\right\rangle_0\cr }\eqno(3.28)$$
where $s=-{1\over \alpha_+}(\sum _{i=1}^N\beta_i+Q)$ and
$\sum_{i=1}^Nk_i+nd_++md_- =2\alpha_0$. After fixing the
$\widehat  {SL_2}$
symmetry as before and integrating over the Grassmann variables
the amplitude
becomes
$$
\eqalignno{
{\cal A}^{nm}_N&=\Gamma(-s)\left( -{\pi\over 2}\right)^3\left(
{i\mu\alpha_+^2\over \pi} \right)^s(-d_+^2)^n(-d_-^2)^m\cr
&\times \prod_{j=4}^N\int d^2\tilde  z_j\prod_{i=1}^n\int d^2t_i
\prod_{i=1}^m\int d^2r_i\prod _{i=1}^sd^2w_i\vert w_i\vert ^{-2\alpha_+
\beta_1}\vert
1-w_i\vert ^{-2\alpha_+\beta_2}\cr
&\times \prod_{i<j}\vert w_i-w_j\vert^{-2\alpha_+^2}
\prod_{i=1}^s\prod_{j=4}^N\vert w_i-\tilde  z_j\vert^{-2\alpha_+
\beta_j} \cr
&\prod_{j=4}^N\vert \tilde  z_j\vert ^{2(k_1k_j-\beta_1\beta_j)}\vert
1-\tilde  z_j\vert ^{2(k_2k_j-\beta_2\beta_j)}\prod _{j<l=4}^N\vert
\tilde  z_j-\tilde  z_l\vert ^{2(k_jk_l-\beta_j\beta_l)}\cr
&\times\prod_{i=1}^n\vert t_i\vert ^{2k_2d_+}\vert 1-t_i\vert^{2k_2d_+}
\prod_{i<j}^n\vert t_i-t_j\vert^{2d_+^2}\prod_{i=1}^n
\prod_{j=1}^m\vert t_i-r_j\vert ^{-2}\cr
&\times\prod_{i=1}^m\vert r_i\vert ^{2k_jd_-}\vert 1-r_i\vert^{2k_2d_-}
\prod_{i<j}^m\vert r_i-r_j\vert^{2d_-^2}\prod_{i=1}^n
\prod_{j=4}^N\vert t_i-\tilde  z_j\vert ^{2d_+k_j}
\prod_{i=1}^m\prod_{j=4}^N\vert r_i-\tilde  z_j\vert ^{2d_-k_j}\cr
&\times \!\left\langle \!(\beta^2_1\overline \psi \psi (0)\!-\!k_1^2
\overline \xi \xi(0))
\prod_{j=4}^N\!(\beta^2_j\overline \psi\psi(\tilde  z_j)\!-\!k_j^2\overline
\xi \xi (\tilde  z_j))\prod_{i=1}^n\!\overline \xi\xi(t_i)\!\prod_{i=1}^m\!
\overline\xi\xi(r_i)\! \prod_{i=1}^s\overline \psi\psi(w_i)\!\right\rangle_0
& (3.29)\cr}
$$
Now we have several terms which give non-trivial amplitudes,
in the following we assume $m+n$ and $N+s$ even; thus we have:
$$
\eqalignno{
{\cal A}^{nm}_N&=\Gamma(-s)\left( -{\pi\over 2}\right)^3\left(
{i\mu\alpha_+^2\over \pi} \right)^s(-d_+^2)^n(-d_-^2)^m \left(
\prod_{j=4}^N\beta^2_j\right) (\alpha_+)^{-2}\cr
&\times \alpha^2\prod_{j=4}^N\int d^2\tilde  z_j\prod_{i=1}^n\int d^2t_i
\prod_{i=1}^m\int d^2r_i
\prod _{i=1}^sd^2w_i\vert w_i\vert ^{2\alpha}\vert
1-w_i\vert ^{2\beta}\cr
&\times \prod_{i<j}\vert w_i-w_j\vert^{4\rho}
\prod_{i=1}^s\prod_{j=4}^N\vert w_i-\tilde  z_j\vert ^{2p_j}
\prod_{j=4}^N\vert \tilde  z_j\vert ^{2\alpha_j}\vert
1-\tilde  z_j\vert ^{2\beta_j}\prod _{j<l=4}^N\vert
\tilde  z_j-\tilde  z_l\vert ^{2\rho_{jl}}\cr
&\times\prod_{i=1}^n\vert t_i\vert ^{2\tilde  \alpha}
\vert 1-t_i\vert^{2\tilde  \beta}
\prod_{i<j}^n\vert t_i-t_j\vert^{2\tilde  \rho}
\prod_{i=1}^n\prod_{j=1}^m\vert t_i-r_j\vert ^{-2}\cr
&\times\prod_{i=1}^m\vert r_i\vert ^{2\tilde  \alpha'}
\vert 1-r_i\vert^{2\tilde\beta'}\prod_{i<j}^m\vert r_i-r_j\vert^{2\tilde\rho'}
\times \prod_{i=1}^n\prod_{j=4}^N\vert t_i-\tilde  z_j\vert ^{2\tilde
\alpha_j}\prod_{i=1}^m\prod_{j=4}^N\vert r_i-\tilde  z_j\vert ^{2\tilde
\alpha'_j }\cr
&\times \left\langle \prod_{i=1}^n\overline \xi \xi (t_i)\prod_{i=1}^m
\overline \xi \xi(r_i)\right\rangle _0\left\langle\overline \psi\psi (0)
\prod_{j=1}^s \overline \psi\psi(w_i)\prod_{j=4}^N\overline
\psi \psi (\tilde  z_j)\right\rangle _0 \quad .&(3.30)\cr}
$$
The definitions of the kinematics parameters are defined as in the bosonic case

We shall use  the kinematics: $k_1,k_2,\cdots,k_{N-1}\ge\alpha_0\, ,
\, k_N<\alpha_0\le 0$ in order to eliminate all parameters
in terms of $\alpha,\beta,\rho$ and $p_j \, (4\le j\le N-1)$.

The symmetries:
$$
{\cal A}_N^{nm}(\alpha,\beta,\rho,p_1,p_2,\cdots, p_{N-1})={\cal
A}_N^{nm}(\beta,\alpha,\rho,p_1,p_2,\cdots, p_{N-1})\eqno(3.31)
$$
$$
{\cal A}_N^{nm}(\alpha,\beta,\rho,p_1\cdots p_{N-1})\! =\!
{\cal A}_N^{nm}(\! - \alpha \! - \! \beta \! + \!
{(m \! - \! 1)\over 2}\!-\! P\! + \rho(N \! +\! n \! - \! s\! -\! 1\!),\beta,
\rho,p_1\cdots  p_{N-1}\!)
$$
with  $P =\sum \limits_{j=4}^{N-1}p_j$ and the large-$\alpha$ behaviour: $
{\cal A}_N^{nm}(\alpha \! \to\! \infty)\! \sim \!
\alpha^{1-m+2\beta+2\rho(s-N-n+3)+2P}$ motivate us to write down the ansatz:
$$
{\cal A}_N^{nm}\! =\! f_N^{nm}\!(\rho,p_1,\cdots ,p_{N-1}\!)
\Delta({1\over
2}\!+\!\rho\!-\!\alpha)\Delta(\!{1\over 2}\!+\!\rho\!-\!\beta\!)
\Delta (\!1-
{m\over 2}\!+\!P\!+\!\alpha\!+\!\beta\!
+\!\rho(\!2\!+\!s\!-\!n\!-\!N\!))
\eqno(3.32)
$$
Taking the limit  $k_i \to 0\, \, (3\le i\le N-1)$,
which implies $p_j\to 2\rho \,
 (4\le j\le N-1)$, we can determine $f_N^{nm}(\rho,p_1,
\cdots p_{N-1})$ using:
$$
{\cal A}_N^{nm}(\alpha,\beta,\rho,k_i\to 0)=\left( -{i\pi\over
2}\right)^{N-3}{\partial ^{N-3}\over \partial _\mu}{\cal A}_3^{nm}
(k_1,k_2,k_N )\eqno(3.33)
$$
and the result for ${\cal A}_3^{nm}$ (see (3.23)). We arrive at the result
$$\eqalignno{
{\cal A}_N^{nm}&=(s+N-3)(s+N-4)\cdots (s+1)\left[ \mu \Delta ({1\over
2}-\rho)\right]^s\cr
&\times\! \left[ -{i\pi\over 2}\Delta({1\over
2}+\rho)\right] ^n\! \left[ -{i\pi\over 2}\Delta({1\over 2}+{\rho^{-1}\over 4})
\right] ^m \!\prod_{i=1}^N \left( -{i\pi\over 2}\right)^N\Delta({1\over
2}(1+\beta_i^2-k_i^2))&(3.34)\cr }
$$
Redefining $\Psi_{NS}\, ,\, \mu $ and
the screening charges we have our final
result:
$$
{\cal A}_N^{nm}={\partial\over \partial\mu}^{N-3}\mu^{s+N-3}\quad ,\eqno(3.35)
$$
which has the same functional form as the bosonic amplitude (2.42). As in that
case the above result is correct only in the kinematic region used to calculate
it and has to be continued outside this region.

\vskip 1cm
\penalty-200
\centerline {\bf 4- Conclusion }
\vskip 1cm
 \nobreak

In  the first part of this paper we have generalized previous results\ref{9}
for the
$N$-point tachyon correlator in Liouville theory coupled to conformal matter
(on the sphere) with $c\le 1$ to the case which includes screening charges in
the matter sector. The results might be useful in understanding the issue of
fusion rules in the calculation of the 3-point correlator (see discussion in
[20]).

In the second part we have obtained the $N$-point NS-correlators in super
Liouville  coupled to $\hat c\le 1$ matter (also on the sphere), including
screening charges thus generalizing the results of [21,26] to the limit case
$\hat c=1$,
 ($N >3$), and the results of [20], obtained for $s=0$, to any value of $s$.

We have obtained, explicit formulae for the corresponding 2D-integrals involved
and the final form of the amplitudes factorizes in the $N$-external legs
factors, confirming the results obtained in [20] (for $s=0$) through a
detailed analysis of the pole structure of the integrals.

In our calculations it was possible to see the singularity in the
renormalization of the cosmological constant ($\mu \to \tilde \mu
/\Delta({1\over 2}-\rho)$) at the point $\hat c=1$ ($\rho=-1/2$). This is
similar to the bosonic case where $\mu \to \tilde \mu/\Delta(-\rho)$ and the
$c=1 \, ,\,  (\rho=-1)$ point is also singular. The final (renormalized)
$N$-point  amplitude has the same form of the  bosonic one. The similarity
with the bosonic case has been found before in the discrete approach [20,25].

Finally we should stress that our results must be continued to other kinematic
regions, this is likely to be very similar to the bosonic case (without s.c.)
worked out in [20].

As a further development it would be interesting to carry out analogous
calculations in the case with $N=2$ supersymmetry and see, among other aspects,
whether the
barrier at $c=3$ indeed disappears. The inclusion of the Ramond sector (for
$s\ne 0$) is also of interest.
\vskip .5cm
 {\bf Acknowledgments}
\vskip .3cm
The work of K.H. (contract \# 90/1799-9) and D.D. (contract \# 90/2246-3) was
supported by FAPESP while the work of E.A. and M.C.B.A. is partially supported
by CNPq.

\vskip .5cm
\penalty-100
\centerline {\bf References}
\vskip .5cm
\nobreak
\refer[[1]/N. Seiberg, Lecture at 1990 Yukawa Int. Sem. Common Trends in Math.
and Quantum Field Theory, and Cargese meeting Random Surfaces, Quantum Gravity
and Strings, May 27, June 2, 1990; J.
Polchinski, Strings '90 Conference, College Station, TX, Mar 12-17, 1990, Nucl.
 Phys. {\bf B357}(1991)241.]

\refer[[2]/E. Br\'ezin and V. A. Kazakov, Phys. Lett.  {\bf B236}
(1990)144; M. R. Douglas and S. H. Shenker, Nucl. Phys.  {\bf B335}(1990)635;
D. J. Gross and A. A. Migdal, Phys. Rev. Lett. {\bf 64}(1990)127.]

\refer[[3]/L. Alvarez-Gaum\'e, H. Itoyama, J.L. Ma\~nes and A. Zadra, Prep.
CERN-TH-6329/91.]

\refer[[4]/A. M. Polyakov, Phys. Lett.  {\bf B103}(1981)207.]

\refer[/A. M. Polyakov, Phys. Lett.  {\bf B103 }(1981)211.]

\refer[[5]/T. L. Curtright and C. B. Thorn, Phys. Rev. Lett. {\bf 48},
(1982)1309; E. Braaten, T. L. Curtright and C. B. Thorn, Phys. Lett.  {\bf
B118}(1982)115; Ann. Phys. {\bf 147}(1983)365; E. Braaten, T. L.
Curtright, G. Gandour and C. B. Thorn, Phys. Rev.
Lett. {\bf 51}(1983)19; Ann. Phys. {\bf 153}(1984)147; J. L. Gervais and A.
Neveu, Nucl. Phys.  {\bf B199}(1982)50;  {\bf B209}(1982)125;  {\bf B224}
(1983)329;  {\bf B238}(1984)123,396; E. D'Hoker and R. Jackiw, Phys.
Rev.  {\bf D26}(1982)3517; T. Yoneya, Phys. Lett.  {\bf B148}(1984)111.]

\refer[[6]/J.-L. Gervais and A. Neveu, Nucl. Phys. {\bf B199} (1982) 59; {\bf
B209} (1982) 125. ]

\refer[/J.-L. Gervais and A. Neveu, Nucl. Phys. {\bf B224} (1983) 329.]

\refer[/J.-L. Gervais and A. Neveu, Phys.  Lett. {\bf B151} (1985) 271.]

\refer[/J.-L. Gervais and A. Neveu, Nucl. Phys. {\bf B238} (1984) 396.]

\refer[[7]/M. Goulian and M. Li, Phys. Rev. Lett. {\bf 66}(1991)2051.]

\refer[[8]/N. Sakai and Y. Tanii, Prog. Theor. Phys. {\bf 86} (1991)547.]

\refer[[9]/P. Di Francesco and D. Kutasov, Phys. Lett.  {\bf B261}(1991)385.]

\refer[[10]/Vl. S. Dotsenko, Mod. Phys. Lett. {\bf A6} (1991)3601.]

\refer[[11]/Y. Kitazawa,  Phys. Lett. {\bf B265} (1991)262.]

\refer[[12]/A. Gupta, S. P. Trivedi and M. B. Wise, Nucl. Phys.  {\bf B340}
(1990)475.]

\refer[[13]/M. Bershadsky and I. R. Klebanov, Phys. Rev. Lett. {\bf 65},
3088 (1990); Nucl. Phys. {\bf B360} (1991)559;
A. M. Polyakov, Mod. Phys. Lett.  {\bf A6}(1991)635.]

\refer[[14]/K. Aoki and E. D'Hoker, preprint UCLA/91/TEP/32(1991).]

\refer[[15]/J. F. Arvis,
Nucl.\ Phys.\  {\bf B212}(1983)151;  {\bf B218}(1983)309; O. Babelon,
Nucl. Phys. {\bf B258}(1985)680; Phys. Lett. {\bf 141B}(1984)353; T.
Curtright and G. Ghandour, Phys. Lett. {\bf B136}(1984)50.]

\refer[[16]/Vl. S. Dotsenko and V. Fateev, Nucl. Phys.  {\bf B240}
(1984)312;  {\bf B251}(1985)691.]

\refer[[17]/J. Distler, Z. Hlousek and H. Kawai, Int. J. Mod. Phys. {\bf A5}
(1990)391.]

\refer[[18]/E. Martinec, Phys. Rev. {\bf D28}(1983)2604.]

\refer[[19]/E. Abdalla, M.C.B. Abdalla, D. Dalmazi and K. Harada,
``Correlation functions in super Liouville theory", Phys. Rev. Lett. (to
appear).]

\refer[[20]/P. Di Francesco and D. Kutasov, Princeton Univ. preprint PUPT-1276
(1991).]

\refer[[21]/K. Aoki and E. D'Hoker, preprint UCLA/91/TEP/33(1991).]

\refer[[22]/F. David, Mod. Phys. Lett. {\bf A3}(1988)1651; J. Distler and
H. Kawai, Nucl. Phys.  {\bf B321}(1989)509.]

\refer[[23]/A.M. Polyakov, Mod. Phys. Lett. {\bf A7} (1991) 635.]

\refer[[24]/E. Abdalla, M.C.B. Abdalla, D. Dalmazi and K. Haradda General NS
correlators in super Liouville theory. IFT-P042/91.]

\refer[[25]/L. Alvarez-Gaum\'e and J. L. Ma\~nez, Mod. Phys. Lett.
{\bf A6}, 2039 (1991).]

\refer[[26]/L. Alvarez-Gaum\'e and Ph. Zaugg, Phys. Lett. {\bf B273} (1991)
81.]

\refer[[27]/Vl. Dotsenko, Paris VI Preprint PAR-LEPTHE 91-52 (1991).]

\refer[[28]/L. Alvarez-Gaum\'e et al., CERN-TH 6142/91 (1991).]

\refer[[29]/M. Bershadsky, V. Knizhnik and A. Teilteman, Phys. Lett.
{\bf B151} (1985) 31.]

\refer[[30]/E. d'Hoker,   Phong, Rev. Mod. Phys. ]


\end